\documentclass[prd,preprint,nofootinbib]{revtex4-1}

\usepackage{color}
\usepackage[latin1]{inputenc}
\usepackage[T1]{fontenc}
\usepackage{amsmath}
\usepackage{amssymb}
\usepackage{amsthm}
\usepackage{mathrsfs}
\usepackage{fancyhdr}
\usepackage{graphicx}
\usepackage{epstopdf}
\usepackage{subfigure}
\usepackage{placeins}
\usepackage{array}
\usepackage[all]{xy}
\usepackage{eufrak}
\usepackage{euscript}
\usepackage{enumerate}
\usepackage{slashed}
\usepackage{hyperref}
\usepackage{hypcap}

\hypersetup{pdftex,colorlinks=true,linkcolor=blue,citecolor=blue,menucolor=black,urlcolor=blue,filecolor=blue}

\begin{document}

\title{Holographic calculation of the QCD crossover temperature in a magnetic field}

\author{Romulo Rougemont}
\email{romulo@if.usp.br}
\affiliation{Instituto de F\'{i}sica, Universidade de S\~{a}o Paulo, C.P. 66318, 05315-970, S\~{a}o Paulo, SP, Brazil}

\author{Renato Critelli}
\email{renato.critelli@usp.br}
\affiliation{Instituto de F\'{i}sica, Universidade de S\~{a}o Paulo, C.P. 66318, 05315-970, S\~{a}o Paulo, SP, Brazil}

\author{Jorge Noronha}
\email{noronha@if.usp.br}
\affiliation{Instituto de F\'{i}sica, Universidade de S\~{a}o Paulo, C.P. 66318, 05315-970, S\~{a}o Paulo, SP, Brazil}
\affiliation{Department of Physics, Columbia University, 538 West 120th Street, New York, NY 10027, USA}

%\date{\today}

\begin{abstract}
Lattice data for the QCD equation of state and the magnetic susceptibility computed near the crossover transition at zero magnetic field are used to determine the input parameters of a five dimensional Einstein-Maxwell-Dilaton holographic model. Once the model parameters are fixed at zero magnetic field, one can use this holographic construction to study the effects of a magnetic field on the equilibrium and transport properties of the quark-gluon plasma. In this paper we use this model to study the dependence of the crossover temperature with an external magnetic field. Our results for the pressure of the plasma and the crossover temperature are in quantitative agreement with current lattice data for values of the magnetic field $0 \le eB \lesssim 0.3$ GeV$^2$, which is the relevant range for ultrarelativistic heavy ion collision applications.\\

\noindent \textbf{Keywords:} Holography, thermodynamics, magnetic field, equation of state, magnetic susceptibility, crossover transition.
\end{abstract}

\maketitle
\tableofcontents

%%%%%%%%%%%%%%%%%%%%%%%%%

\section{Introduction}
\label{introduction}

Recent relativistic heavy ion collision experiments \cite{expQGP1,expQGP2,expQGP3,expQGP4} have produced a strongly coupled quark-gluon plasma (QGP) \cite{QGP} whose physical properties are currently under intense investigation (see \cite{reviewQGP1,reviewQGP2} for recent reviews). The study of the equilibrium and transport properties of the QGP as functions of parameters such as the temperature $T$, chemical potential(s), and (electro)magnetic fields are of great relevance for the characterization and understanding of this new state of QCD matter. In particular, very strong magnetic fields up to $\mathcal{O}\left(0.3\,\textrm{GeV}^2\right)$ are expected to be created in the early stages of noncentral relativistic heavy ion collisions \cite{noncentralB1,noncentralB2,noncentralB3,noncentralB4,noncentralB5,noncentralB6}\footnote{It is not clear at the moment if the electromagnetic fields present in the early stages of heavy ion collisions remain strong enough to directly affect equilibrium and transport properties of the plasma produced at later stages. However, see \cite{Gursoy:2014aka} for a recent study on how the QGP's electric conductivity may actually delay the decay of the magnetic field in the medium.} and even much larger magnetic fields of $\mathcal{O}\left(4\,\textrm{GeV}^2\right)$ may have been produced in the early stages of the Universe \cite{universe1,universe2} (see also Fig.\ 10 in \cite{latticedata0}). Moreover, magnetic fields up to $\mathcal{O}\left(1\,\textrm{MeV}^2\right)$ are present in the interior of very dense neutron stars known as magnetars \cite{magnetar}. Therefore, the study of the effects of strong magnetic fields on the QGP has sparked a large amount of interest in the community in recent years \cite{Agasian:2008tb,Mizher:2010zb,Evans:2010xs,Preis:2010cq,Fukushima:2012xw,Fukushima:2012kc,Bali:2012zg,Blaizot:2012sd,Callebaut:2013ria,Bali:2013esa,Bonati:2014ksa,Fukushima:2013zga,Machado:2013rta,Fraga:2013ova,Andersen:2013swa,Bali:2013owa,Ferreira:2013oda,Ruggieri:2014bqa,Ferreira:2014kpa,Farias:2014eca,Ayala:2014iba,Ayala:2014gwa,Ferrer:2014qka,Kamikado:2014bua,Yu:2014xoa,Braun:2014fua,Mueller:2015fka,Endrodi:2015oba} (for extensive reviews and other references, see for instance, \cite{Fraga:2012rr,reviewfiniteB1,reviewfiniteB2,reviewfiniteB3}).

Since the properties of a strongly coupled QGP cannot be reliably studied using perturbative techniques one has to resort to nonperturbative approaches that are valid at strong coupling. Interestingly enough, contrary to what happens in the case of a nonzero baryon chemical potential where the sign problem of the fermion determinant prevents the application of the Monte Carlo importance sampling method in lattice simulations (for a review see \cite{fodorreview}), in the case of a nonzero magnetic field (at vanishing baryon chemical potential) standard lattice techniques may be employed to study the equilibrium properties of QCD in the $(T,B)$-plane, see for instance, \cite{latticedata0,latticedata2,latticedata3}.
 
Another nonperturbative method that is suited to study strongly coupled non-Abelian gauge theories is the holographic anti-de Sitter/conformal field theory (AdS/CFT) correspondence (also known as the gauge/gravity duality) \cite{adscft1,adscft2,adscft3}. The correspondence has been employed to obtain useful insights into the properties of the strongly coupled QGP, as recently reviewed in \cite{solana,adams}. A very attractive feature of the gauge/gravity duality is that it may be easily employed to compute transport coefficients of strongly coupled non-Abelian gauge theory plasmas (see, for instance, \cite{GN2,conductivity,hydro,gubser2,finitemu,baryondiff,Finazzo:2015xwa}), which is a challenging task to perform on the lattice \cite{Meyer:2011gj}.

A top-down holographic dual for $\mathcal{N}=4$ super Yang-Mills theory (SYM) in the presence of an external constant magnetic field was proposed in \cite{DK1,DK2,DK3} and calculations for different physical observables in this scenario were carried out, for instance, in \cite{DK-applications1,DK-applications2,DK-applications3,DK-applications4}. However, the QGP formed in heavy ion collisions \cite{reviewQGP1,reviewQGP2} probes the temperature region within which the QCD plasma is highly nonconformal \cite{latticedata1} (when $T \sim 150-300$ MeV). Therefore, in order to make contact with realistic heavy ion collision applications, one needs to develop holographic models that are able to capture some of the relevant aspects of the physics of the strongly coupled QGP near the QCD crossover \cite{Aoki:2006we}. One possible way to accomplish this within holography is to deform the boundary quantum field theory by turning on a dynamical scalar field in the bulk whose boundary value sources a relevant operator in the gauge theory. Near the boundary the scalar field approaches zero and conformal invariance is recovered in the ultraviolet. In the infrared, however, the holographic dual gauge theory generated by such deformation behaves very differently than a conformal plasma and may be tuned to display some of the properties of QCD in the strong coupling regime.

In this work we construct a nonconformal anisotropic bottom-up holographic model that is suited for the study of a QCD-like plasma at nonzero magnetic field and vanishing chemical potential(s). Our model is built up on classical nonconformal anisotropic black brane solutions to the Einstein-Maxwell-Dilaton (EMD) model defined with a negative cosmological constant and in the presence of an external constant magnetic field. This constitutes a sequel to the studies of strongly coupled nonconformal plasmas via black brane solutions initiated by \cite{GN1,GN2} in the case of finite temperature, zero magnetic field, and vanishing chemical potential\footnote{These nonconformal solutions can also be adapted to study the vacuum properties of the gauge theory, as recently discussed in \cite{stefanovacuum}. This was studied in detail earlier in \cite{ihqcd-1,ihqcd-2,hot-ihqcd} in the case of similar bottom-up models at zero and finite temperature concerning pure glue Yang-Mills theory.}, which was later extended in \cite{gubser1,gubser2} and also \cite{finitemu} to take into account the presence of a nonzero baryon chemical potential at zero magnetic field\footnote{See also \cite{ihqcd-veneziano} for a bottom-up holographic model at finite temperature, nonzero chemical potential, and zero magnetic field in the Veneziano limit \cite{veneziano}.}. This type of nonconformal model has been used in the last years to investigate how different observables of phenomenological relevance to the QGP and the physics of heavy ion collisions vary near the QCD crossover transition. In fact, after the original calculations in \cite{GN1,GN2}, which included the evaluation of the bulk viscosity at zero baryon chemical potential and zero magnetic field \cite{GN2}, a series of other quantities were computed within this type of holographic model such as the heavy quark free energy \cite{Noronha:2009ud,Noronha:2010hb}, the energy loss of highly energetic probes \cite{Ficnar:2010rn,Ficnar:2011yj,Ficnar:2012yu}, the Debye screening mass \cite{stefanovacuum}, the electric conductivity \cite{conductivity}, a large set of first and second order viscous hydrodynamic transport coefficients \cite{hydro}, the spectrum of quasinormal modes \cite{Janik:2015waa} and the thermal photon production rate \cite{yang-muller}. In the context of the holographic models developed in \cite{gubser1,gubser2} and \cite{finitemu} as extensions of the original models \cite{GN1,GN2}, taking into account the presence of a nonvanishing baryon chemical potential, we mention the calculation of the holographic critical point in the $(T,\mu_B)$-plane and the associated critical exponents \cite{gubser1}, the evaluation of the holographic equation of state, the heavy quark drag force, the Langevin diffusion coefficients, the jet quenching parameter, the energy loss of light quarks and an estimate of the equilibration time in the baryon-rich strongly coupled QGP \cite{finitemu}, the evaluation of the bulk viscosity \cite{gubser2}, as well as the baryon susceptibility, baryon conductivity, thermal conductivity, baryon diffusion \cite{baryondiff}, and the thermal photon and dilepton production rates \cite{Finazzo:2015xwa} at finite baryon chemical potential and zero magnetic field. Here we add one more entry to this family of nonconformal black hole solutions by taking into account, for the first time, the presence of a magnetic field in the nonconformal, QCD-like gauge theory.

Our model is a bottom-up holographic setup in which the dilaton potential and the Maxwell-Dilaton gauge coupling are dynamically fixed in order to describe lattice data at zero chemical potential(s) and vanishing magnetic field, which should be contrasted with top-down models coming from compactifications of known string theory solutions. Although in bottom-up models the holographic dual is not precisely known, the fact that these models may be constructed using some phenomenological input from QCD makes it possible that at least part of the physics of the boundary gauge field theory resembles, even at the quantitative level, QCD in the strong coupling limit. Thus, one may regard such constructions as holographic effective theories that are engineered to model some specific aspects of QCD phenomenology. Once the model parameters are fixed, these theories can be used to make predictions about observables that are currently beyond the scope of lattice calculations, such as most of the second order hydrodynamic coefficients \cite{hydro}.

This paper is organized as follows. In Section \ref{sec2} we describe in detail the construction of our holographic model and how the dilaton potential and the Maxwell-Dilaton gauge coupling can be determined by lattice data for the $(2+1)$-flavor lattice QCD equation of state and magnetic susceptibility at zero magnetic field, respectively. With the holographic model parameters fully specified, we proceed in Section \ref{sec3} to obtain the holographic equation of state at nonzero magnetic field and present results for the temperature and magnetic field dependence of the entropy density and the pressure. We find that the deconfinement temperature in our holographic model decreases with an increasing magnetic field, as recently observed on the lattice. Moreover, our model results for the pressure and the crossover temperature are in quantitative agreement with current lattice data up to $eB \lesssim 0.3$ GeV$^2$, which is the relevant range of magnetic fields for heavy ion collisions. We present our conclusions in Section \ref{conclusion} where we also point out other applications to be pursued in the near future using the anisotropic nonconformal holographic model developed here.

Throughout this paper we use natural units $c=\hbar=k_B=1$ and a mostly plus metric signature.

\section{The holographic model}
\label{sec2}

Assuming as usual that charm quarks are not relevant in the crossover transition, in QCD there are three different chemical potentials associated with three independent globally conserved charges. These different chemical potentials are the three lighter quark chemical potentials $\mu_u$, $\mu_d$, $\mu_s$ or, equivalently, the baryon chemical potential $\mu_B$, the electric charge chemical potential $\mu_Q$, and the strangeness chemical potential $\mu_S$. For each nonzero chemical potential in the gauge theory there must be a nonzero temporal component of the associated gauge field in the bulk. It is also clear that an Abelian magnetic field $B$ in the gauge theory should come from a nonzero spatial component of the gauge potential in the electric charge sector.\footnote{See, for instance, \cite{DK3} for a construction with finite $B$ and finite electric charge density in the context of Einstein-Maxwell-Chern-Simons theories.}

In the present work we solely focus on the electric charge sector at $B \neq 0$ with $\mu_Q=\mu_B=\mu_S=0$, which may be described by the following EMD action
\begin{align}
S&=\frac{1}{16\pi G_5}\int_{\mathcal{M}_5}d^5x\sqrt{-g}\left[R-\frac{1}{2}(\partial_\mu\phi)^2-V(\phi) -\frac{f(\phi)}{4}F_{\mu\nu}^2\right] +S_{\textrm{GHY}}+S_{\textrm{CT}},
\label{2.1}
\end{align}
where $S_{\textrm{GHY}}$ is the Gibbons-Hawking-York action \cite{ghy1,ghy2} needed to establish a well-posed variational problem with Dirichlet boundary condition for the metric, and $S_{\textrm{CT}}$ is the counterterm action that can be constructed using the holographic renormalization procedure \cite{ren1,ren2,ren3,ren4,ren5}. These two boundary terms contribute to the total on-shell action but not to the equations of motion and, since we shall not need to compute the total on-shell action in the present work, we do not need to worry about their explicit form here. Also, as we are going to discuss in detail in Section \ref{sec2.4}, we shall dynamically fix the gravitational constant $G_5$, the dilaton potential $V(\phi)$, and the Maxwell-Dilaton gauge coupling $f(\phi)$, by solving the equations of motion for the EMD fields with the requirement that the holographic equation of state and magnetic susceptibility at zero magnetic field match the corresponding lattice QCD results.

In \eqref{2.1}, the metric field in the bulk is dual to the stress-energy tensor of the boundary field theory while the dilaton field is introduced in order to dynamically break the conformal symmetry of the gauge theory in the infrared. The Abelian gauge field in the bulk is employed here to introduce an external magnetic field at the boundary, which we take to be constant and uniform in the $\hat{z}$-direction and, as stated before, in the present work we set all the chemical potentials to zero. The constant and uniform magnetic field breaks the $SO(3)$ rotational invariance of the gauge theory down to $SO(2)$ rotations around the $\hat{z}$-axis implying that the Ansatz for the bulk metric must be anisotropic and translationally invariant. Also, at zero temperature this Ansatz must be invariant under boosts in the $(t,z)$-plane though this symmetry is not present at nonzero temperature. Based on these symmetry properties, which are phenomenologically dictated by the corresponding symmetry content present in current lattice QCD calculations defined on the $(T,B)$-plane, we take the following black brane Ansatz for the bulk fields in\footnote{As we shall discuss soon, $\mathcal{B}$ is one of the two initial conditions controlling the temperature and the external magnetic field at the boundary quantum field theory. The other initial condition corresponds to the value of the dilaton field evaluated at the black brane horizon, $\phi_0$. The set of initial conditions $(\phi_0,\mathcal{B})$ is nontrivially related to the thermodynamical pair $(T,B)$ in the gauge theory. In Sections \ref{sec2.3} and \ref{sec2.4} we discuss how one can relate $\mathcal{B}$ to the external magnetic field at the boundary gauge theory, $B$.} \eqref{2.1}:
\begin{align}
ds^2&=e^{2a(r)}\left[-h(r)dt^2+dz^2\right]+e^{2c(r)}(dx^2+dy^2)+\frac{e^{2b(r)}dr^2}{h(r)},\nonumber\\
\phi&=\phi(r),\,\,\,A=A_\mu dx^\mu=\mathcal{B}xdy\Rightarrow F=dA=\mathcal{B}dx\wedge dy,
\label{2.2}
\end{align}
where the radial location of the black brane horizon, $r_H$, is given by the largest root of the equation $h(r_H)=0$ and in our coordinates the boundary of the asymptotically AdS$_5$ spacetime is located at $r\rightarrow\infty$. In \eqref{2.2} we have already fixed a convenient gauge for the Maxwell field, which in the present case is a prescribed non-dynamical field. Also, for simplicity, we shall adopt units where the asymptotic AdS$_5$ radius is equal to one.

Using \eqref{2.2}, the equations of motion obtained from \eqref{2.1} may be expressed as follows
\begin{align}
\phi''+\left(2a'+2c'-b'+\frac{h'}{h}\right)\phi'-\frac{e^{2b}}{h} \left(\frac{\partial V(\phi)}{\partial\phi}+\frac{\mathcal{B}^2e^{-4c}}{2}\frac{\partial f(\phi)}{\partial\phi}\right)&=0,\label{2.3}\\
a''+\left(\frac{14}{3}c'-b'+\frac{4}{3}\frac{h'}{h}\right)a' +\frac{8}{3}a'^2+\frac{2}{3}c'^2+\frac{2}{3}\frac{h'}{h}c'
+\frac{2}{3}\frac{e^{2b}}{h} V(\phi)-\frac{1}{6}\phi'^2&=0,\label{2.4}\\
c''-\left(\frac{10}{3}a'+b'+\frac{1}{3}\frac{h'}{h}\right)c' +\frac{2}{3}c'^2-\frac{4}{3}a'^2-\frac{2}{3}\frac{h'}{h}a'
-\frac{1}{3}\frac{e^{2b}}{h} V(\phi)+\frac{1}{3}\phi'^2&=0,\label{2.5}\\
h''+\left(2a'+2c'-b'\right)h'&=0,\label{2.6}
\end{align}
where the prime denotes a derivative with respect to the radial direction. Using these equations of motions one can also derive a useful constraint
\begin{align}
a'^2+c'^2-\frac{1}{4}\phi'^2+\left(\frac{a'}{2}+c'\right)\frac{h'}{h}+4a'c'
+\frac{e^{2b}}{2h}\left(V(\phi)+\frac{\mathcal{B}^2e^{-4c}}{2}f(\phi)\right)=0.
\label{2.7}
\end{align}
The equation of motion for the Maxwell field is automatically satisfied by the Ansatz \eqref{2.2}. Moreover, $b(r)$ has no equation of motion and, thus, it can be freely chosen to take any value due to reparametrization invariance. In the next Section we specify a subsidiary condition for $b(r)$ that defines a convenient gauge for the metric that will be used in the numerical calculations carried out in the present work.

\subsection{Ultraviolet expansions}
\label{sec2.1}

For the calculation of physical observables in the gauge theory one needs to obtain the near-boundary, far from the horizon expansions for the bulk fields $a(r)$, $c(r)$, $h(r)$, and $\phi(r)$. In the present work, we use the domain-wall gauge defined by the subsidiary condition $b(r)=0$. At the boundary the dilaton field goes to zero in such a way that $V(\phi(r\rightarrow\infty)\rightarrow 0)=-12$ and $f(0)$ is a finite positive constant\footnote{Note that in \eqref{2.1} the Maxwell-Dilaton gauge coupling $f(\phi)$ plays the role of an inverse effective gauge coupling squared and, therefore, it must correspond to a positive-definite function.}. Also, the metric blackening factor, $h(r)$, must go to a constant at the boundary, which we denote by\footnote{This constant is equal to one in the so-called ``standard coordinates'' of the domain-wall gauge, which we shall discuss soon. Here we are considering general coordinates where this constant may be different than one. We shall also see later how to relate these two sets of coordinates.} $h(r\rightarrow\infty)=h_0^{\textrm{far}}$.

Moreover, since we are interested in asymptotically AdS$_5$ solutions to the equations of motion \eqref{2.3}, \eqref{2.4}, \eqref{2.5}, and \eqref{2.6}, at the boundary one finds $a(r\rightarrow\infty)=c(r\rightarrow\infty)$. In the domain-wall gauge $b(r)=0$, the leading order near-boundary expression for $a(r)$ (and also $c(r)$) is linear in $r$ \cite{gubser1,gubser2} such that at lowest order in $\phi(r\rightarrow\infty)\rightarrow 0$ we may consider the following leading order far from the horizon ultraviolet asymptotics
\begin{align}
V(\phi)\approx -12,\,\,\, f(\phi)\approx f(0),\,\,\,h(r)\approx h_0^{\textrm{far}},\,\,\,
a(r)\approx a_0^{\textrm{far}}+a_{-1}^{\textrm{far}}r,\,\,\, c(r)\approx c_0^{\textrm{far}}+c_{-1}^{\textrm{far}}r,
\label{2.8}
\end{align}
where $a_{-1}^{\textrm{far}}=c_{-1}^{\textrm{far}}$, as discussed above. Indeed, by substituting \eqref{2.8} into the equations of motion and taking the asymptotic limit of large $r$ (where the ultaviolet expansions hold), one concludes that
\begin{align}
a_{-1}^{\textrm{far}}=c_{-1}^{\textrm{far}}=\frac{1}{\sqrt{h_0^{\textrm{far}}}}.
\label{2.9}
\end{align}
In order to obtain the next to leading order term for $h(r)$ and also the first terms for $\phi(r)$ in the ultraviolet expansions for the bulk fields, we consider the first backreaction of the near-boundary fields expressed in \eqref{2.8} and \eqref{2.9} on the equations of motion\footnote{This procedure may be repeated to obtain all the other subleading terms in the ultraviolet expansions. However, we only need the first few terms in these expansions to compute the thermodynamical observables.}. In fact, we first consider the next to leading order near-boundary expansion for the dilaton potential
\begin{align}
V(\phi)\approx -12+\frac{m^2}{2}\phi^2,\,\,\, m^2=-\nu\Delta,
\label{2.10}
\end{align}
where $\Delta$ is the ultraviolet scaling dimension of the gauge invariant operator dual to the bulk dilaton field and we defined $\nu=d-\Delta$, where $d=4$ is the dimension of the boundary. We shall see in Section \ref{sec2.4} that a good description of lattice data can be achieved  by taking $\Delta\approx 3$ ($\nu\approx 1$). One can now show that the far from horizon ultraviolet asymptotics for the bulk fields may be written as
\begin{align}
a(r)&\approx\alpha(r)+\cdots,\nonumber\\
c(r)&\approx\alpha(r)+(c_0^{\textrm{far}}-a_0^{\textrm{far}})+\cdots,\nonumber\\
h(r)&\approx h_0^{\textrm{far}}+h_4^{\textrm{far}}e^{-4\alpha(r)}+\cdots,\nonumber\\
\phi(r)&\approx \phi_Ae^{-\nu\alpha(r)}+\phi_Be^{-\Delta\alpha(r)}+\cdots,
\label{2.11}
\end{align}
where we defined $\alpha(r)=a_0^{\textrm{far}}+r/\sqrt{h_0^{\textrm{far}}}$ while $\cdots$ denotes subleading terms. We note that the ultraviolet asymptotics \eqref{2.11} are in agreement with our numerical solutions. By comparing these numerical solutions to \eqref{2.11} one can determine the ultraviolet coefficients $a_0^{\textrm{far}}$, $c_0^{\textrm{far}}$, $h_0^{\textrm{far}}$ and $\phi_A$, which are needed to compute the thermodynamical observables in Sections \ref{sec2.3} and \ref{sec2.4}.

\subsection{Infrared expansions}
\label{sec2.2}

Now we consider the infrared, near-horizon expansions for the bulk fields $a(r)$, $c(r)$, $h(r)$, and $\phi(r)$. Near the horizon all the bulk fields in \eqref{2.2} are assumed to be smooth and we may consider the Taylor expansions
\begin{align}
X(r)=\sum_{n=0}^\infty X_n(r-r_H)^n,
\label{2.12}
\end{align}
where $X=\left\{a,c,h,\phi\right\}$.

In order to numerically solve the equations of motion \eqref{2.3}, \eqref{2.4}, \eqref{2.5}, and \eqref{2.6} we need to specify the boundary conditions $X(r_{\textrm{start}})$ and $X'(r_{\textrm{start}})$, where $r_{\textrm{start}}$ is a value of the radial coordinate that is slightly above the horizon\footnote{The horizon is a singular point of the equations of motion and, thus, we need to initialize the numerical integrations slightly above it.}. In this paper we work with Taylor expansions up to second order, which are sufficient to perform the numerical integrations if $r_{\textrm{start}}$ is close enough to $r_H$. Therefore, we must determine 12 Taylor coefficients in order to specify $X(r_{\textrm{start}})$ and $X'(r_{\textrm{start}})$ at second order. One of these 12 coefficients, namely, $\phi_0$, is one of the two initial conditions of the problem\footnote{As discussed before, the other initial condition is $\mathcal{B}$.}. Four of these 12 coefficients, namely, $a_0$, $c_0$, $h_0$, and $h_1$ and also the radial location of the black hole horizon, $r_H$, may be fixed by rescaling the bulk coordinates while taking into account also the fact that $h(r)$ vanishes at the horizon. For definiteness, we adopt here numerical coordinates fixed in such a way that
\begin{align}
r_H=0;\,\,\,a_0=c_0=h_0=0,\,\,\,h_1=1.
\label{2.13}
\end{align}
Note that $r_H=0$ may be obtained by rescaling the radial coordinate while $h_0=0$ comes from the fact that $h(r)$ has a simples zero at the horizon. Also, $h_1=1$ may be obtained by rescaling $t$ while $a_0=0$ may be arranged by rescaling $(t,z)$ by a common factor. Similarly, $c_0=0$ may be arranged by rescaling $(x,y)$ by a common factor. After this, the remaining 7 coefficients in the near-horizon Taylor expansions for the bulk fields can be fixed on-shell as functions of the initial conditions $(\phi_0,\mathcal{B})$ by substituting the second order Taylor expansions into the equations of motion and setting to zero each power of $r_{\textrm{start}}$ in the resulting algebraic equations\footnote{In practice, we set to zero the following 7 terms: $\mathcal{O}(r_{\textrm{start}}^{0})$, $\mathcal{O}(r_{\textrm{start}}^{1})$, and $\mathcal{O}(r_{\textrm{start}}^{2})$ in \eqref{2.6}, $\mathcal{O}(r_{\textrm{start}}^{-1})$ in \eqref{2.7}, $\mathcal{O}(r_{\textrm{start}}^{-1})$ and $\mathcal{O}(r_{\textrm{start}}^{0})$ in \eqref{2.3}, and $\mathcal{O}(r_{\textrm{start}}^{0})$ in \eqref{2.4}.}.

With $X(r_{\textrm{start}})$ and $X'(r_{\textrm{start}})$ determined as discussed above, the equations of motion are numerically integrated from $r_{\textrm{start}}$ near the horizon up to some numerical ultraviolet cutoff $r_{\textrm{max}}$ near the boundary. We used $r_{\textrm{start}}=10^{-8}$ and $r_{\textrm{max}}=10$ to numerically solve the equations of motion. It is important to remark, however, that even before reaching $r_{\textrm{conformal}}=2$ the numerical backgrounds we considered in the present work have already reached the ultraviolet fixed point corresponding to the AdS$_5$ geometry. This fact is used in Section \ref{sec2.3} to reliably obtain the ultraviolet coefficients in \eqref{2.11} and it will be also employed in Section \ref{sec2.4} to properly compute the holographic magnetic susceptibility numerically.

\subsection{Coordinate transformations and thermodynamical observables}
\label{sec2.3}

Let us now introduce the so-called ``standard coordinates'' of the domain-wall metric gauge, $\tilde{b}(\tilde{r})=0$, where variables with $\sim$ refer to quantities evaluated in these standard coordinates where the background reads
\begin{align}
d\tilde{s}^2&=e^{2\tilde{a}(\tilde{r})}\left[-\tilde{h}(\tilde{r})d\tilde{t}^2+d\tilde{z}^2\right]+ e^{2\tilde{c}(\tilde{r})}(d\tilde{x}^2+d\tilde{y}^2)+\frac{d\tilde{r}^2}{\tilde{h}(\tilde{r})},\nonumber\\
\tilde{\phi}&=\tilde{\phi}(\tilde{r}),\,\,\, \tilde{A}=\tilde{A}_\mu d\tilde{x}^\mu=\hat{B}\tilde{x}d\tilde{y}\Rightarrow \tilde{F}=d\tilde{A}=\hat{B}d\tilde{x}\wedge d\tilde{y},
\label{2.14}
\end{align}
and the boundary is at $\tilde{r}\rightarrow\infty$ while the horizon is at $\tilde{r}=\tilde{r}_H$. The ``hat'' in $\hat{B}$ accounts for the fact that this is the magnetic field measured in units of the inverse of the AdS radius squared, while $B$ shall be used to denote the boundary magnetic field measured in physical units, as we shall discuss in Section \ref{sec2.4}. In the standard coordinates, the ultraviolet asymptotics for the bulk fields are given by \cite{gubser1,gubser2} (see also \cite{finitemu})
\begin{align}
\tilde{a}(\tilde{r})&\approx\tilde{r}+\cdots,\nonumber\\
\tilde{c}(\tilde{r})&\approx\tilde{r}+\cdots,\nonumber\\
\tilde{h}(\tilde{r})&\approx 1+\cdots,\nonumber\\
\tilde{\phi}(\tilde{r})&\approx e^{-\nu\tilde{r}}+\cdots.
\label{2.15}
\end{align}

The standard coordinates (in which $h(r)$ goes to one at the boundary) are the coordinates where we obtain standard holographic formulas for the gauge theory's physical observables such as the temperature and the entropy density. However, in order to obtain numerical solutions for the bulk fields one needs to give numerical values for all the infrared near-horizon Taylor expansion coefficients, which in turn requires rescaling these standard coordinates, as discussed in the previous Section. The numerical solutions are obtained in the numerical coordinates described by the Ansatz \eqref{2.2} with the ultraviolet asymptotics \eqref{2.11}, while standard holographic formulas for physical observables are obtained in the standard coordinates described by the background \eqref{2.14} with the ultraviolet asymptotics \eqref{2.15}. One may relate these two sets of coordinates by equating $\tilde{\phi}(\tilde{r})=\phi(r)$, $d\tilde{s}^2=ds^2$ and $\hat{B}d\tilde{x}\wedge d\tilde{y}=\mathcal{B}dx\wedge dy$ and this leads to the following relations\footnote{As mentioned in \cite{gubser1}, if $\phi_A<0$ one must replace $\phi_A\mapsto|\phi_A|$ in these relations.} (by comparing the near-boundary asymptotics \eqref{2.11} and \eqref{2.15} for $r\rightarrow\infty$)
\begin{align}
\tilde{r}&=\frac{r}{\sqrt{h_0^{\textrm{far}}}}+a_0^{\textrm{far}}-\ln\left(\phi_A^{1/\nu}\right),\nonumber\\
\tilde{t}&=\phi_A^{1/\nu}\sqrt{h_0^{\textrm{far}}}t,\nonumber\\
\tilde{x}&=\phi_A^{1/\nu}e^{c_0^{\textrm{far}}-a_0^{\textrm{far}}}x,\nonumber\\
\tilde{y}&=\phi_A^{1/\nu}e^{c_0^{\textrm{far}}-a_0^{\textrm{far}}}y,\nonumber\\
\tilde{z}&=\phi_A^{1/\nu}z;\nonumber\\
\tilde{a}(\tilde{r})&=a(r)-\ln\left(\phi_A^{1/\nu}\right),\nonumber\\
\tilde{c}(\tilde{r})&=c(r)-(c_0^{\textrm{far}}-a_0^{\textrm{far}})-\ln\left(\phi_A^{1/\nu}\right),\nonumber\\
\tilde{h}(\tilde{r})&=\frac{h(r)}{h_0^{\textrm{far}}},\nonumber\\
\tilde{\phi}(\tilde{r})&=\phi(r);\nonumber\\
\hat{B}&=\frac{e^{2(a_0^{\textrm{far}}-c_0^{\textrm{far}})}}{\phi_A^{2/\nu}}\mathcal{B}.
\label{2.16}
\end{align}

The temperature of the plasma is given by the black brane horizon's Hawking temperature
\begin{align}
\hat{T}=\frac{\sqrt{-\tilde{g}'_{\tilde{t}\tilde{t}} \tilde{g}^{\tilde{r}\tilde{r}}\,'}}{4\pi}\biggr|_{\tilde{r}=\tilde{r}_H}= \frac{e^{\tilde{a}(\tilde{r}_H)}}{4\pi}|\tilde{h}'(\tilde{r}_H)|=\frac{1}{4\pi\phi_A^{1/\nu}\sqrt{h_0^{\textrm{far}}}},
\label{2.17}
\end{align}
while the entropy density is obtained via the Bekenstein-Hawking's relation \cite{bek1,bek2}
\begin{align}
\hat{s}=\frac{S}{V}=\frac{A_H/4G_5}{V}=\frac{\int_{\textrm{horizon}}d^3\tilde{x}\sqrt{\tilde{g}(\tilde{r}=\tilde{r}_H, \,\tilde{t}\,\textrm{fixed})}}{4G_5V} = \frac{2\pi}{\kappa^2}e^{\tilde{a}(\tilde{r}_H)+2\tilde{c}(\tilde{r}_H)} = \frac{2\pi e^{2(a_0^{\textrm{far}}-c_0^{\textrm{far}})}}{\kappa^2\phi_A^{3/\nu}},
\label{2.18}
\end{align}
where we defined $\kappa^2=8\pi G_5$ and used \eqref{2.12}, \eqref{2.13}, and \eqref{2.16}.

One can see from \eqref{2.16}, \eqref{2.17}, and \eqref{2.18} that the only ultraviolet coefficients in the numerical coordinates which we need to fix by fitting the numerical solutions with \eqref{2.11} are $a_0^{\textrm{far}}$, $c_0^{\textrm{far}}$, $h_0^{\textrm{far}}$, and $\phi_A$. The numerical solutions for $h(r)$ converge quickly to their asymptotic values at large $r$ and we may reliably set $h_0^{\textrm{far}}=h(r_{\textrm{conformal}})$. With $h_0^{\textrm{far}}$ fixed in this way, we may fix $a_0^{\textrm{far}}$, $c_0^{\textrm{far}}$, and $\phi_A$, respectively, by employing the fitting functions $a(r)=a_0^{\textrm{far}}+r/\sqrt{h_0^{\textrm{far}}}$, $c(r)=c_0^{\textrm{far}}+r/\sqrt{h_0^{\textrm{far}}}$, and $\phi(r)=\phi_Ae^{-\nu a(r)}$ in the interval $r\in[r_{\textrm{conformal}}-1,r_{\textrm{conformal}}]$. We were able to obtain good fits for the near-boundary behavior of the numerical solutions using this fitting scheme.

\begin{figure}[h]
\begin{centering}
\includegraphics[scale=0.55]{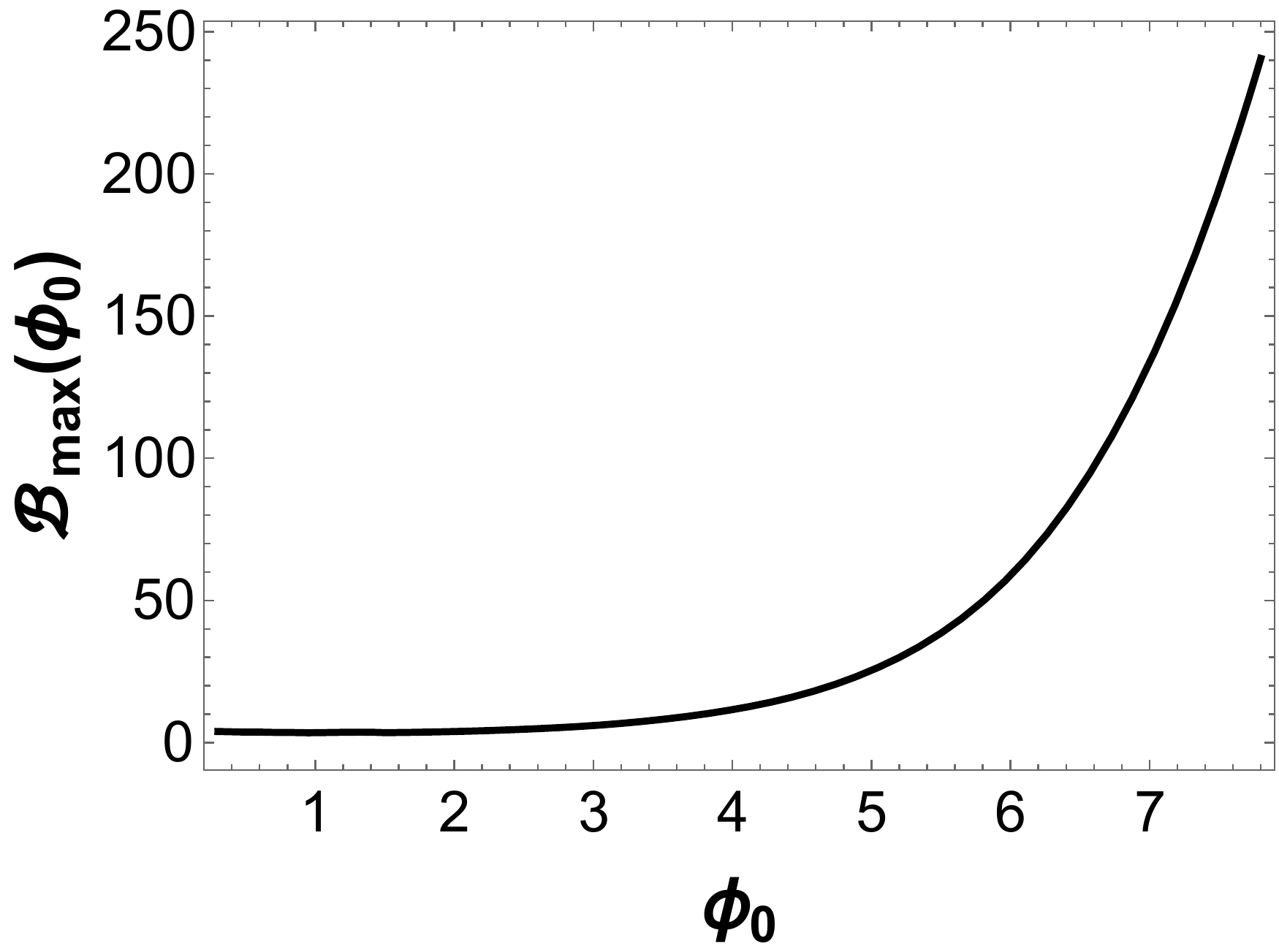}
\par\end{centering}
\caption{The curve corresponds to the upper bound for the initial condition $\mathcal{B}$ as a function of the initial condition for the dilaton $\phi_0$, below which the solutions of the equations of motion are asymptotically AdS$_5$. This curve depends on the chosen profiles for the dilaton potential $V(\phi)$ and gauge coupling function $f(\phi)$ to be discussed in the next Section. \label{fig1}}
\end{figure}

Also, it is important to remark that there is an upper bound on the initial condition $\mathcal{B}$ for a given value of the initial condition for the dilaton $\phi_0$. In fact, for values of $\mathcal{B}$ above this bound, all the numerical backgrounds we generated failed to be asymptotically AdS$_5$. Such a bound, which we denote by $\mathcal{B}\le\mathcal{B}_{\textrm{max}}(\phi_0)$, may be numerically constructed by interpolating a list with pairs of points $\left\{\left(\phi_0^i,\mathcal{B}_{\textrm{max}}^i\right),\,i=1,2,3,\cdots\right\}$ and the corresponding result is presented in Fig.\ \ref{fig1}. 

In the next Section we explain how one can express the thermodynamical quantities $\hat{B}$, $\hat{T}$, and $\hat{s}$ in physical units\footnote{Note from \eqref{2.16}, \eqref{2.17}, and \eqref{2.18} that $\hat{B}$, $\hat{T}$, and $\hat{s}$ are proportional to $\phi_A^{-2/\nu}$, $\phi_A^{-1/\nu}$ and $\phi_A^{-3/\nu}$, respectively. Correspondingly, their counterparts in physical units (without the ``hat'') are given in MeV$^2$, MeV, and MeV$^3$, respectively. This is related to the fact that the leading mode for the dilaton field, $\phi_A$, corresponds to the insertion of a relevant deformation in the quantum field theory, which is responsible for generating an infrared scale that breaks the conformal invariance of the theory at low energies \cite{gubser2}.} using the lattice data for the equation of state and the magnetic susceptibility at zero magnetic field.

\subsection{Fixing the Maxwell-Dilaton gauge coupling using lattice data for the magnetic susceptibility at zero magnetic field}
\label{sec2.4}

Refs.\ \cite{hydro,finitemu} discussed in detail how to dynamically fix the dilaton potential, $V(\phi)$, and the gravitational constant, $\kappa^2$, using the recent lattice data \cite{latticedata1} for the QCD equation of state with $(2+1)$-flavors. We refer the reader to those papers for the details about this procedure. The results are
\begin{align}
V(\phi)=-12\cosh(0.606\,\phi)+0.703\,\phi^2-0.1\,\phi^4+0.0034\,\phi^6,\,\,\,\kappa^2 = 8\pi G_5 = 12.5\,.
\label{2.19}
\end{align}
From the dilaton potential specified above one obtains the dilaton mass $m^2\approx-3$, as anticipated in Section \ref{sec2.1}.

We remark that, although the present EMD construction do not explicit introduce fundamental flavors at the dual boundary quantum field theory, the dilaton potential in Eq. \eqref{2.19} was adjusted in order to quantitatively \emph{mimic} the $(2+1)$-flavor lattice QCD equation of state and its crossover. This \emph{mimicking} procedure was originally introduced in \cite{GN1} (see also \cite{Yaresko:2015ysa} for more recent discussions), where it was also discussed how different choices for the dilaton potential may \emph{emulate} not only the QCD crossover, as done in the present work, but also first and second order phase transitions, which may be useful for a large variety of different physical systems.

\begin{figure}[h]
\begin{tabular}{cc}
\includegraphics[width=0.48\textwidth]{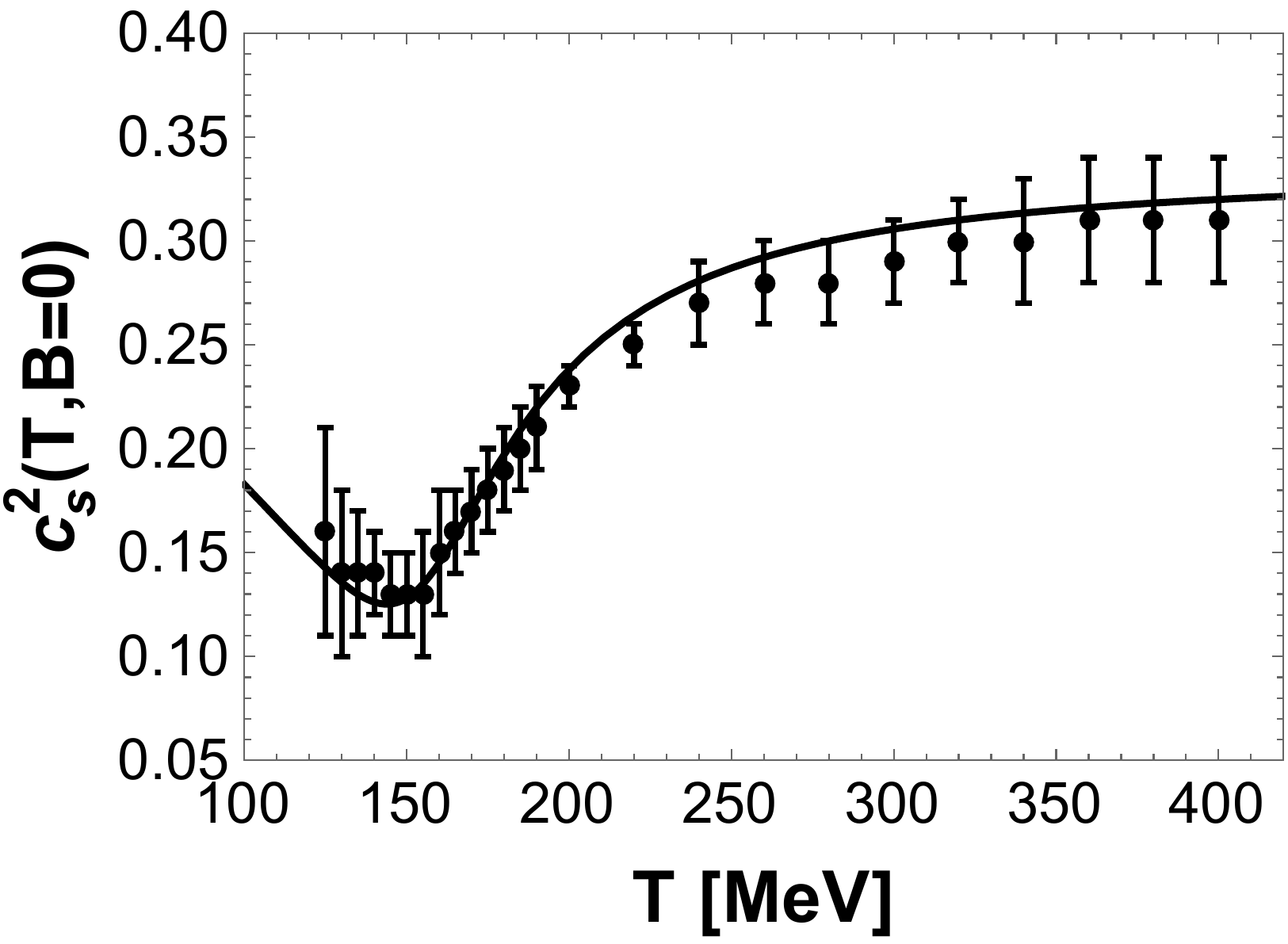} & \includegraphics[width=0.46\textwidth]{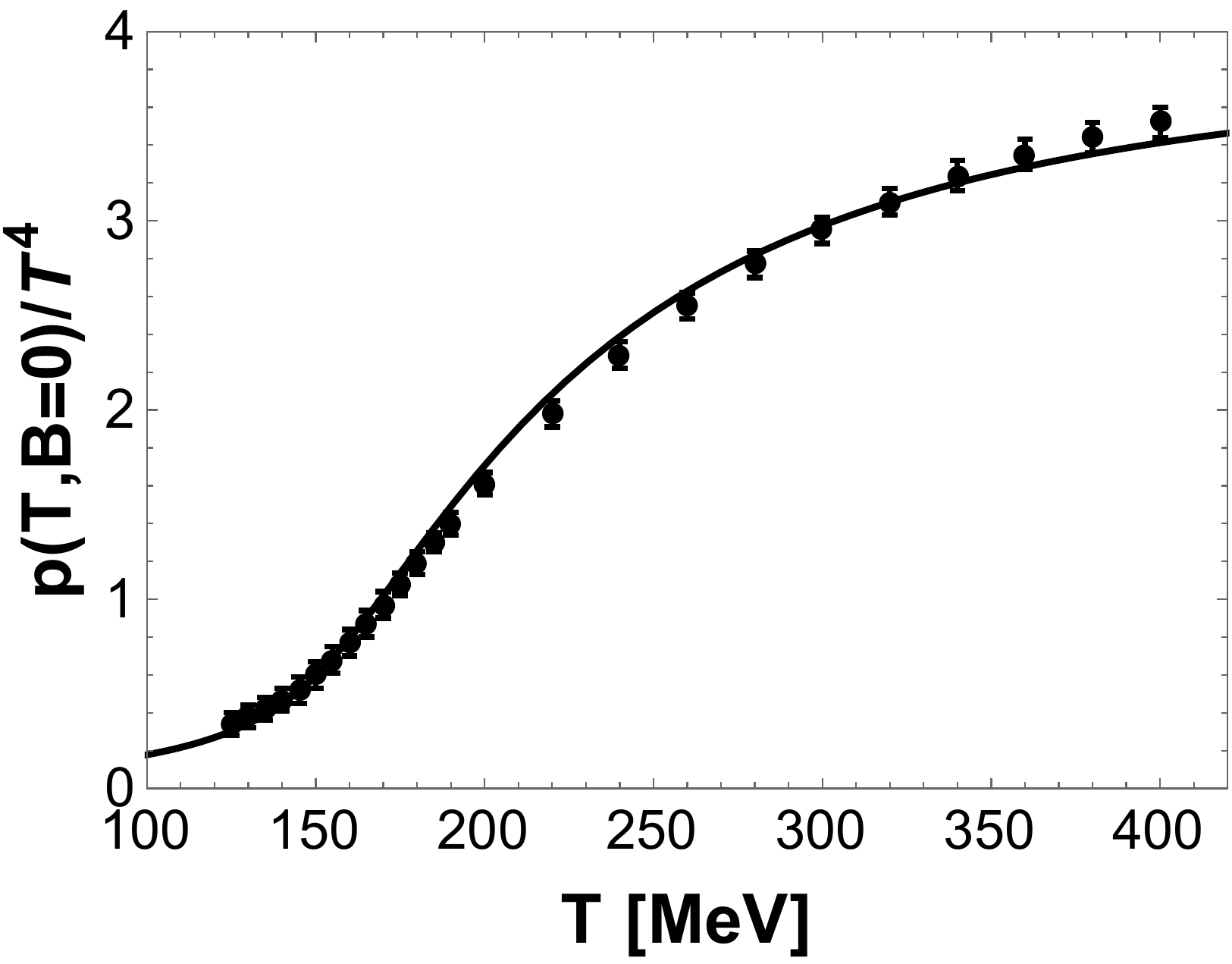} % \\
\end{tabular}
\caption{Holographic calculation of the speed of sound squared $c_s^2$ and the (normalized) pressure $p/T^4$. The data points correspond to lattice QCD results from \cite{latticedata1} computed at zero magnetic field. \label{fig2}}
\end{figure}

In the present paper, we employ the same procedure used in \cite{hydro,finitemu} to express the holographically determined thermodynamical observables in physical units, i.e., we find the temperature at which our speed of sound squared, $c_s^2$, displays a minimum (at zero magnetic field) and match it to the corresponding lattice QCD result \cite{latticedata1}
\begin{align}
\lambda=\frac{T_{\textrm{min.}\,c_s^2}^{\textrm{lattice}}}{T_{\textrm{min.}\,c_s^2}^{\textrm{BH}}}\approx \frac{143.8\,\textrm{MeV}}{0.173} \approx 831\,\textrm{MeV}.
\label{2.20}
\end{align}
In what follows, we relate any black hole thermodynamical observable, $\hat{X}$, with its counterpart in physical units, $X$, with mass dimension [MeV$^p$], by taking $X = \lambda^p \hat{X}$ [MeV$^p$]. This prescription respects the fact that dimensionless ratios, such as $s/T^3$, must give the same result regardless of the units. A comparison between our holographic results for the speed of sound squared, $c_s^2(T,B=0)$, and the (normalized) pressure, $p(T,B=0)/T^4$ (at zero magnetic field) and the corresponding lattice QCD results from \cite{latticedata1} is shown in Fig.\ \ref{fig2}. One can see that the holographic model provides a good description of the lattice data in the absence of an external magnetic field.  

In order to fully determine our holographic model and include the effects from a magnetic field we also need to dynamically fix the Maxwell-Dilaton gauge coupling $f(\phi)$. This can be done using the recent lattice data \cite{latticedata2} for the magnetic susceptibility of QCD with $(2+1)$-flavors evaluated at zero magnetic field. In order to compute the magnetic susceptibility in our holographic model we follow the same general steps discussed in \cite{donos}: we substitute the Ansatz \eqref{2.2} into the action \eqref{2.1} and calculate the second derivative of the on-shell action with respect to the magnetic field, dividing the result by the entire spacetime volume of the boundary. In order to obtain the bare magnetic susceptibility we plug the on-shell numerical solutions into the expression obtained in the previous step\footnote{As mentioned in footnote 7 of \cite{donos}, the Euclidean action has the opposite sign of the Lorentzian action.},
\begin{align}
\chi_{\textrm{bare}}(T,B)=-\frac{\partial^2 f_{\textrm{bare}}}{\partial B^2}=-\frac{1}{V_{\textrm{bdy}}}\frac{\partial^2 S_{E,\,\textrm{bare}}^{\textrm{on-shell}}[B]}{\partial B^2} &=
\frac{1}{V_{\textrm{bdy}}}\frac{\partial^2 S_{\textrm{bare}}^{\textrm{on-shell}}[B]}{\partial B^2}\nonumber\\
& = -\frac{1}{2\kappa^2}\int_{\tilde{r}_H}^{\tilde{r}^{\textrm{fixed}}_{\textrm{max}}}d\tilde{r} f(\tilde{\phi}(\tilde{r})) e^{2(\tilde{a}(\tilde{r})-\tilde{c}(\tilde{r}))}\biggr|^{\textrm{on-shell}},
\label{2.21}
\end{align}
where $f_{\textrm{bare}}$ is the bare free energy density and, formally, one should take the limit $\tilde{r}^{\textrm{fixed}}_{\textrm{max}}\rightarrow\infty$. However, in numerical calculations, $\tilde{r}^{\textrm{fixed}}_{\textrm{max}}$ must be a fixed ultraviolet cutoff for all the geometries in order to ensure that the ultraviolet divergence in \eqref{2.21} is independent of the temperature. Since we are interested here in calculating the magnetic susceptibility at zero magnetic field where $a(r)=c(r)$, one obtains from \eqref{2.21}
\begin{align}
\chi_{\textrm{bare}}(T,B=0)= -\frac{1}{2\kappa^2}\int_{\tilde{r}_H}^{\tilde{r}^{\textrm{fixed}}_{\textrm{max}}}d\tilde{r} f(\tilde{\phi}(\tilde{r}))\biggr|^{\textrm{on-shell}}.
\label{2.22}
\end{align}
In order to regularize \eqref{2.22} we follow the same procedure adopted on the lattice \cite{latticedata2} and subtract from \eqref{2.22} the vacuum contribution at zero temperature. Clearly, this removes the ultraviolet divergences since those are temperature independent. More precisely, we subtract the geometry corresponding to $(T_{\textrm{small}},B)\approx(0.005\,\textrm{MeV},0)$, which is generated by the initial conditions $(\phi_0,\mathcal{B})=(7.8,0)$; this is the asymptotically AdS$_5$ geometry with the lowest temperature and zero magnetic field which we could reach in our numerical computations\footnote{Note that $\phi_0=7.8$ corresponds to the local minimum of our dilaton potential \eqref{2.19}. For $\phi_0>7.8$, our dilaton potential becomes non-monotonic and, in practice, we took $\phi_0=7.8$ as the upper bound for the initial condition $\phi_0$ in our numerical calculations to avoid complications with extra singular points in the equations of motion.}. Therefore, we obtain the following holographic formula for the magnetic susceptibility at zero magnetic field (which is valid for any EMD model of the kind considered here)
\begin{align}
\chi(T,B=0)&=\chi_{\textrm{bare}}(T,B=0)-\chi_{\textrm{bare}}(T_{\textrm{small}},B=0)\nonumber\\
&=-\frac{1}{2\kappa^2}\left[\left(\int_{\tilde{r}_H}^{\tilde{r}^{\textrm{fixed}}_{\textrm{max}}}d\tilde{r} f(\tilde{\phi}(\tilde{r}))\right)\biggr|_{T,B=0}-\left(\textrm{same}\right)\biggr|_{T_{\textrm{small}},B=0} \right]^{\textrm{on-shell}}
\nonumber\\
&=-\frac{1}{2\kappa^2}\left[\left(\frac{1}{\sqrt{h_0^{\textrm{far}}}}\int_{r_{\textrm{start}}}^{r^{\textrm{var}}_{\textrm{max}}} dr f(\phi(r))\right)\biggr|_{T,B=0}-\left(\textrm{same}\right)\biggr|_{T_{\textrm{small}},B=0} \right]^{\textrm{on-shell}},
\label{2.23}
\end{align}
where $\tilde{r}^{\textrm{fixed}}_{\textrm{max}}$ must be chosen in such a way that the upper limits of integration in the numerical coordinates satisfy $r_{\textrm{conformal}}\le r^{\textrm{var}}_{\textrm{max}}=\sqrt{h_0^{\textrm{far}}}\left[\tilde{r}^{\textrm{fixed}}_{\textrm{max}}- a_0^{\textrm{far}}+\ln\left(\phi_A^{1/\nu}\right)\right]\le r_{\textrm{max}}$ for all the geometries considered. We found that for $\tilde{r}^{\textrm{fixed}}_{\textrm{max}} \sim 33$ such requirement is met. We also checked that one can vary the value of the ultraviolet cutoff  $\tilde{r}^{\textrm{fixed}}_{\textrm{max}}$ and the results for the holographic magnetic susceptibility do not change, which confirms the stability of our numerical procedure.

\begin{figure}[h]
\begin{centering}
\includegraphics[scale=0.55]{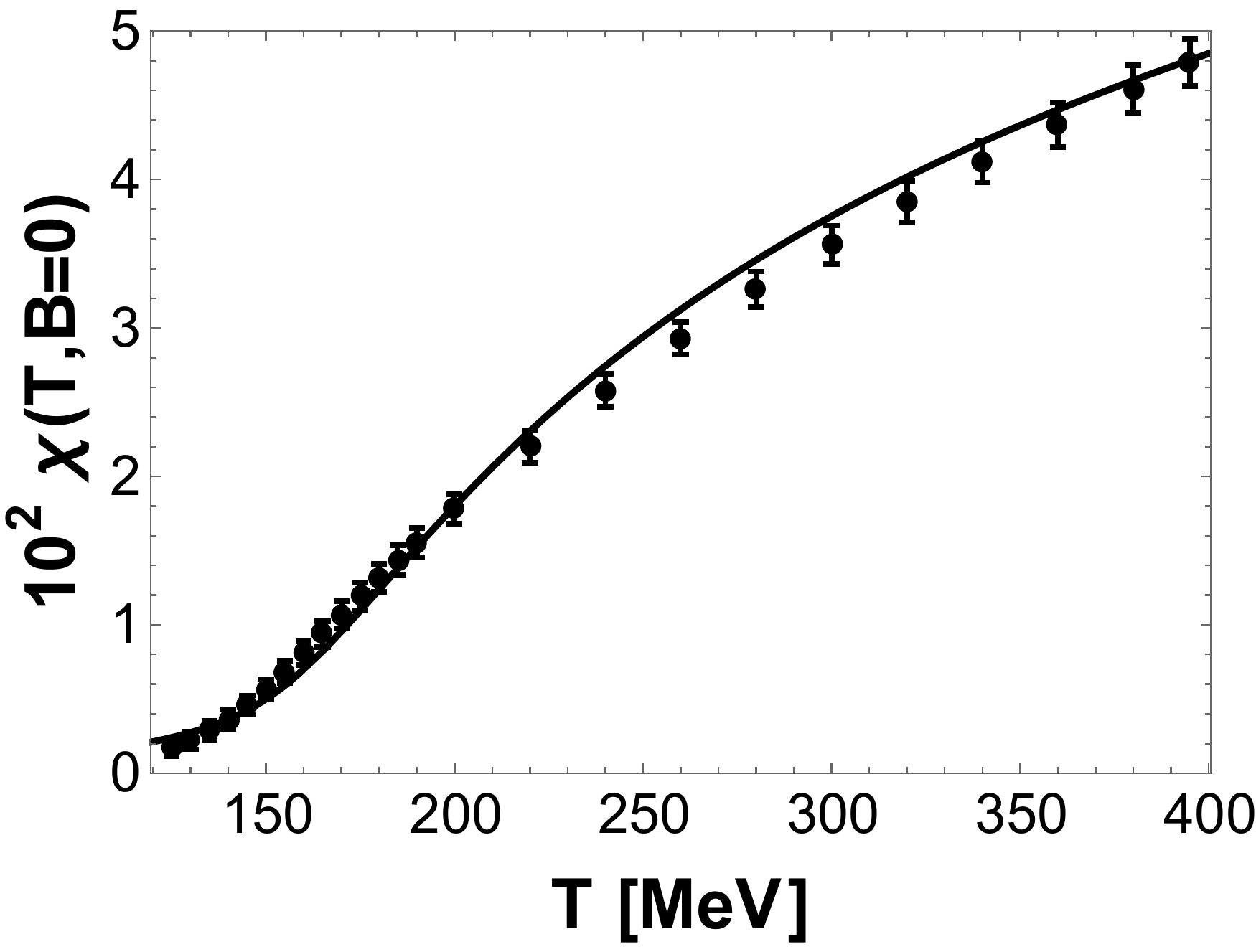}
\par\end{centering}
\caption{Holographic calculation of the magnetic susceptibility at zero magnetic field and comparison with lattice data from \cite{latticedata2} (we consider 10.9 times the data available in table III in \cite{latticedata2}, which corresponds to the magnetic susceptibility in natural units - see footnote 1 in \cite{latticedata2}). \label{fig3}}
\end{figure}

We can now use many different trial profiles for $f(\phi)$ to evaluate \eqref{2.23} over the zero magnetic field background solutions, trying to holographically fit the recent lattice data from \cite{latticedata2} for the magnetic susceptibility of $(2+1)$-flavor QCD at zero magnetic field. We found that a good description of the lattice data can be obtained by fixing
\begin{align}
f(\phi)=1.12\,\textrm{sech}(1.05\,\phi-1.45),
\label{2.24}
\end{align}
with the corresponding results displayed in Fig.\ \ref{fig3}.

With the dilaton potential \eqref{2.19} and the Maxwell-Dilaton gauge coupling \eqref{2.24} dynamically fixed by the description of adequate lattice data at zero magnetic field, our holographic model is now fully determined. This setup may be employed to investigate the physics of the dual quantum field theory at finite temperature and nonzero magnetic field with vanishing chemical potential(s). 

We finish this Section by mentioning some limitations of the holographic model presented here:
\begin{itemize}
\item The model cannot describe phenomena directly related to chiral symmetry and its breaking/restoration (such as $T=0$ magnetic catalysis \cite{Gusynin:1994re,Gusynin:1995nb,Miransky:2002rp}). This could be studied by adding flavor D-branes in the bulk (see, for instance, Ref. \cite{ihqcd-veneziano});\\
\item The model cannot properly describe hadron thermodynamics (which sets in at low temperatures, below $T \sim 150 $ MeV) and the effects of magnetic fields at low temperatures (for a study of the hadron resonance gas in a magnetic field see \cite{Endrodi:2013cs}). Moreover, in this holographic model asymptotic freedom is replaced by conformal invariance at sufficiently high temperatures. Furthermore, for high enough magnetic fields the nonlinear nature of the DBI action for the D-branes should be taken into account \cite{vacilao};\\
\item In Appendix \ref{apa}, we present a brief discussion on the behavior of electric field response functions in the present EMD model, which indicates that this simple model is not versatile enough to simultaneously cover in a quantitative way both the magnetic and electric sectors of the QGP.
\end{itemize}
With these limitations in sight, we expect that the present bottom-up holographic model will be mostly useful to understand the effects of magnetic fields on the QGP within the range $T \sim 150-400$ MeV and $eB\lesssim 1$ GeV$^2$.

\section{Holographic QCD thermodynamics at nonzero magnetic field}
\label{sec3}

\begin{figure}[h]
\begin{tabular}{cc}
\includegraphics[width=0.48\textwidth]{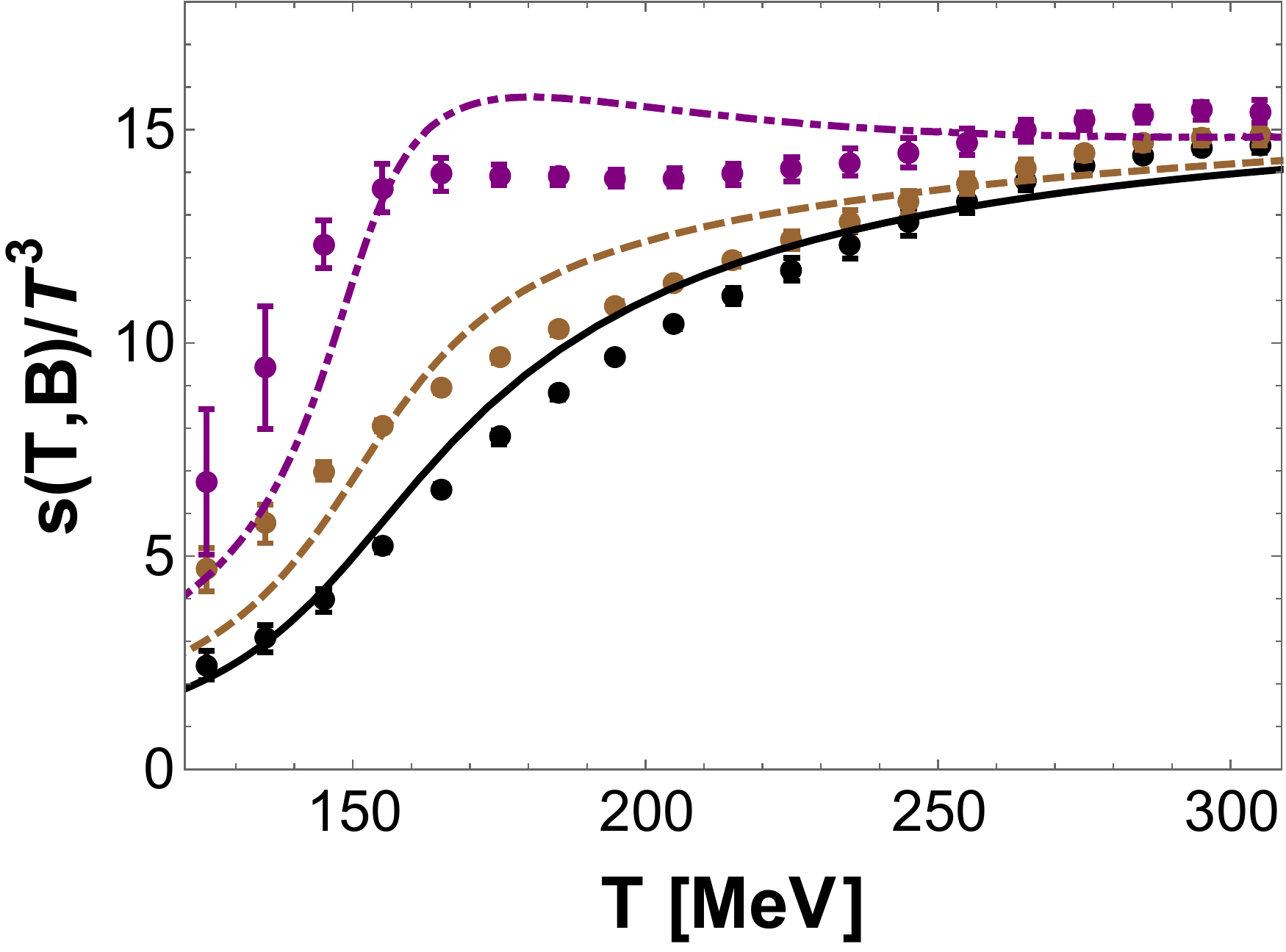} & \includegraphics[width=0.48\textwidth]{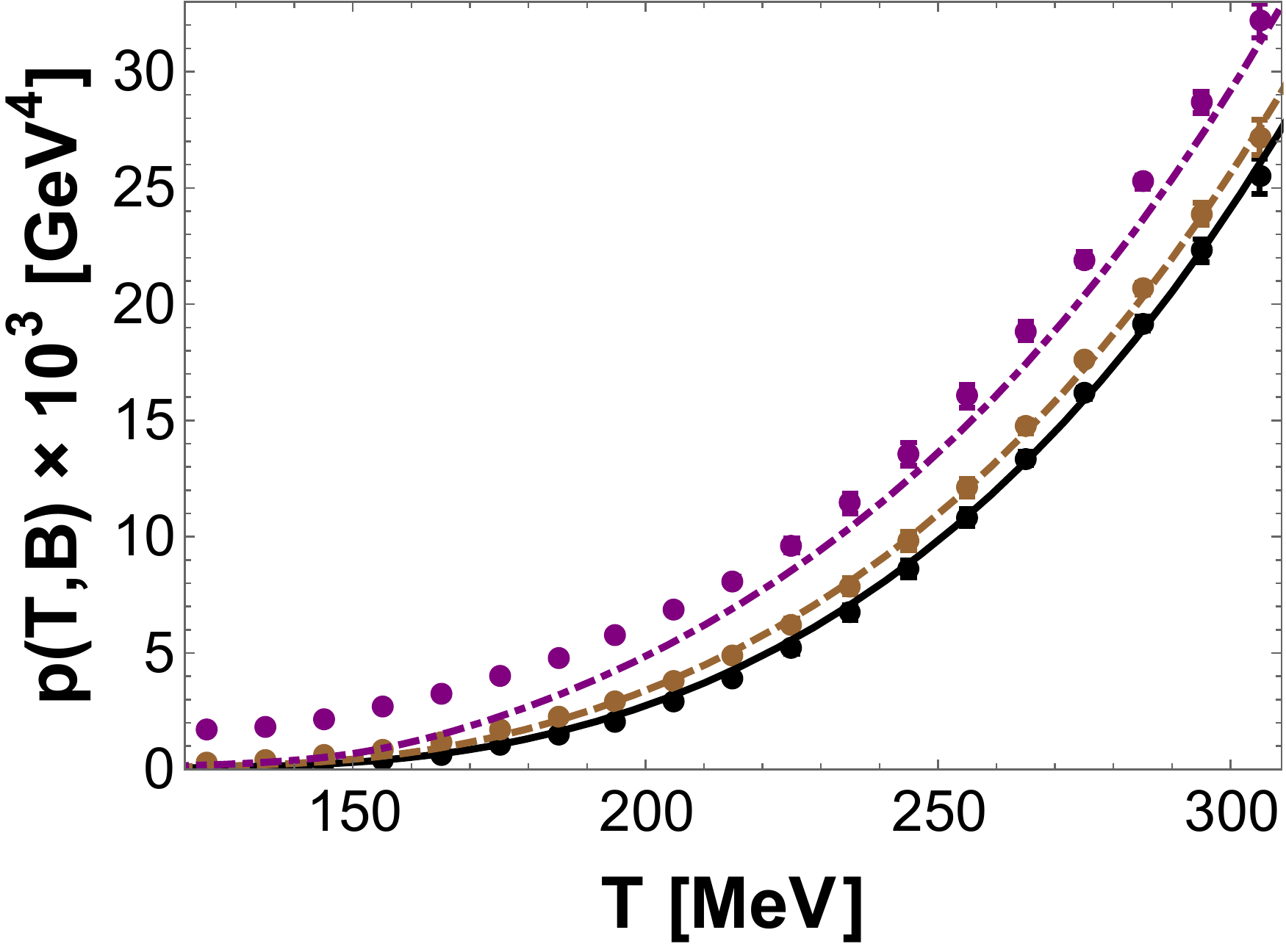} % \\
\end{tabular}
\caption{(Color online) Holographic calculation for the normalized entropy density, $s/T^3$, and pressure, $p$, in the presence of an external magnetic field. The solid, dashed, and dot-dashed curves correspond to magnetic fields $eB=0$, $0.3$, and $0.6$ GeV$^2$, respectively. The data points correspond to the lattice calculations for these quantities performed in \cite{latticedata3}.}
\label{fig4}
\end{figure}

In this Section the results for the holographic equation of state at nonzero magnetic field are presented. The formulas needed to compute the observables shown below were presented in the last Section. Here, we define the pressure as the temperature integral of the entropy density performed while keeping the magnetic field fixed\footnote{As discussed in detail in Section 2 of \cite{latticedata3} this corresponds to the isotropic pressure in the so-called ``$B$-scheme'' where the magnetic field is kept fixed during compression. Also, this corresponds to the anisotropic pressure in the direction of the magnetic field in the so-called ``$\Phi$-scheme'' where the magnetic flux is kept fixed during compression.}
\begin{align}
p(T,B)=\int_{T_{\textrm{ref}}}^T dT' s(T',B),
\label{3.1}
\end{align}
where we took a low reference temperature, $T_{\textrm{ref}}=22$ MeV, in agreement with what was done in \cite{hydro,finitemu} to obtain the fit for the dilaton potential and the gravitational constant \eqref{2.19}. By doing so, the holographic curves for the pressure in Fig. \ref{fig4} (and also Fig. \ref{fig2}) actually correspond to differences with respect to reference pressures calculated at $T_{\textrm{ref}}$ for each value of the magnetic field.

In Fig.\ \ref{fig4} we show our holographic results for the normalized entropy density, $s/T^3$, and pressure, $p$, and compare them to recent lattice data \cite{latticedata3} for $eB = 0$, $0.3$, and 0.6 GeV$^2$. It is important to remark, however, that the above convention to calculate the pressure is not exactly the same used in \cite{latticedata3} since in \eqref{3.1} the pressure (difference) vanishes at $T=T_{\textrm{ref}}=22$ MeV while in the calculation carried out in \cite{latticedata3} the pressure goes like $\sim\mathcal{O}\left((eB)^4\right)$ for $T\rightarrow 0$ and, therefore, one should expect that the differences between these two calculations\footnote{Note that in \cite{latticedata3} the pressure was obtained from the renormalized free energy density. Here, we could have done the analogous holographic procedure by calculating the free energy density from the holographically renormalized on-shell action for the EMD model. This is, however, a much more laborious calculation than the one we have carried out here where we first calculated the entropy density using the Bekenstein-Hawking's relation \eqref{2.18} and then we calculated the pressure (difference) using Eq. \eqref{3.1}.} become more pronounced at low temperatures and large magnetic fields, as seen in Fig.\ \ref{fig4}. However, even for $eB=0.6$ GeV$^2$, we do find a reasonable agreement for the pressure at large temperatures ($T > 200$ MeV).

On the other hand, when it comes to the ratio $s/T^3$, the agreement between our holographic results and the lattice is only at the qualitative level. This is in part due to the uncertainties in the holographic description of this observable already at $B=0$: the holographic model parameters were chosen to describe the lattice data for the pressure and the speed of sound squared at $B=0$ and not\footnote{Probably a better agreement with $B\neq 0$ lattice data may be obtained by improving the choice of the model parameters through a global fit to $B=0$ lattice data for the pressure, the entropy density, the speed of sound, and the trace anomaly.} $s/T^3$. In any case, one can see that $s/T^3$ increases with an increasing magnetic field, which is the general behavior observed on the lattice \cite{latticedata3}. Moreover, note that the curve $s/T^3$ becomes steeper near the transition region for increasing values of the magnetic field, which is again in agreement with the general trend observed on the lattice \cite{Endrodi:2015oba}.

\begin{table}[h]
 \begin{center}
  \begin{tabular}{| c || c |}
    \hline
    $eB$ $[$GeV$^2]$ & $T_c(eB)$ $[$MeV$]$ \\
    \hline
    \hline
    0   & 158.2 \\
    \hline
    0.1 & 157.6 \\
    \hline
    0.2 & 154.9 \\
    \hline
    0.3 & 153.2 \\
    \hline
    0.4 & 151.3 \\
    \hline
    0.5 & 149.9 \\
    \hline
  \end{tabular}
 \caption{Deconfinement temperature (defined by the inflection point of $s/T^3$) for different values of the magnetic field in the bottom-up holographic model. \label{tab1}}
 \end{center}
\end{table}

\begin{figure}[h]
\begin{centering}
\includegraphics[scale=0.55]{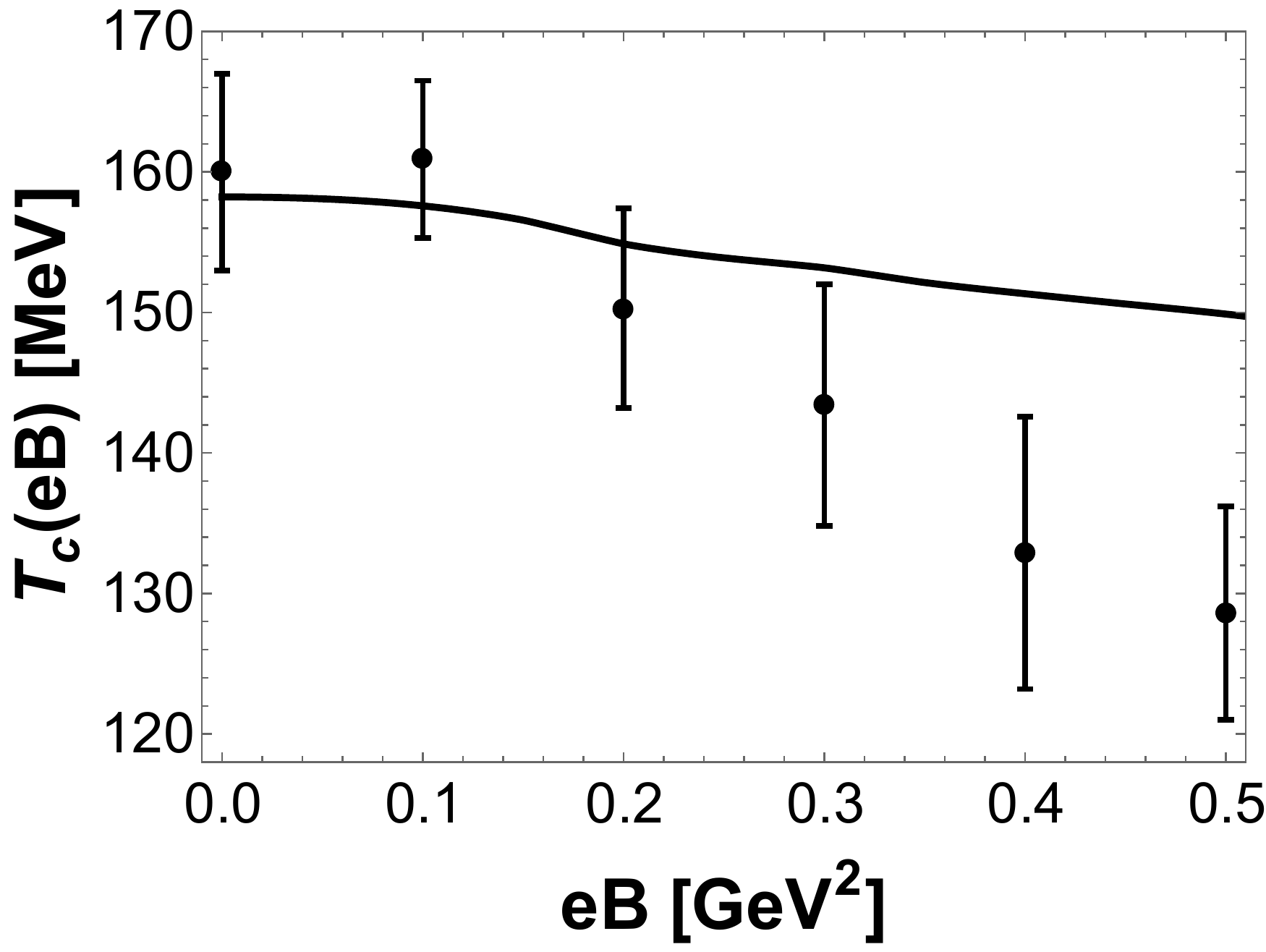}
\par\end{centering}
\caption{Deconfinement temperature (defined by the inflection point of $s/T^3$) for different values of the magnetic field in the bottom-up holographic model. The data points correspond to the lattice calculation performed in \cite{latticedata3}.}
\label{TcB}
\end{figure}

As discussed in \cite{latticedata3}, the inflection point of $s/T^3$ may be used to characterize the crossover temperature as a function of the magnetic field\footnote{Since the crossover is not a genuine phase transition, the free energy is analytic in the region where the degrees of freedom change from a hadron gas to a deconfined plasma. Thus, the definition of the crossover temperature $T_c$ depends on the observable one uses to characterize it. Different observables can give in principle different values for $T_c$ and one may use them to obtain a band defining the crossover region \cite{latticedata0,latticedata3}.}. Correspondingly, the peak in $T\partial_T(s/T^3)$ may be used to estimate the crossover temperature as a function of the magnetic field in our holographic model. We used our results for $s/T^3$ to find how the crossover temperature changes with a magnetic field and the results are displayed in table \ref{tab1} and in Fig.\ \ref{TcB}. One can see in Fig.\ \ref{TcB} that in our model the crossover temperature decreases with an increasing magnetic field, as found on the lattice \cite{latticedata0,latticedata3}, but a quantitative agreement with the data from \cite{latticedata3} occurs only for $eB \lesssim 0.3$ GeV$^2$.

Some general comments regarding the crossover found in our holographic model are in order at this point. Depending on the chosen dilaton potential, the black hole solutions may or may not have a minimum temperature, as detailed discussed, for instance, in Refs. \cite{GN1,stefanovacuum,Yaresko:2015ysa,Charmousis:2010zz}. In the case there is some minimum temperature below which the black hole solutions do not exist, the system generally features a first order Hawking-Page phase transition \cite{Hawking:1982dh} to the thermal gas phase at some critical temperature a little bit higher than the minimum temperature for the existence of the black hole solutions. Also, in this case, the black hole solutions are not unique and there is at least one unstable branch of black hole solutions above this minimum temperature. But for some choices of the dilaton potential the temperature of the black hole solutions may monotonically decrease as a function of the radial position of the horizon until going to zero, in which case the black hole solutions are unique and thermodynamically preferred over the thermal gas solution and the system does not feature any phase transition at nonzero temperature (at least at zero magnetic field and vanishing chemical potentials): this is the case realized in our EMD model. Note also this is indeed the adequate situation to mimic the QCD crossover instead of the pure Yang-Mills first order phase transition. In fact, by analyzing our dilaton potential according to the general criteria discussed in \cite{Charmousis:2010zz}, one notes that in the deep infrared our dilaton potential goes like $V(\phi\to\infty)\sim -e^{0.606\phi}$, in which case at each finite value of temperature (at zero magnetic field and vanishing chemical potentials) there exists a unique black hole solution and this corresponds to the true ground state of the system, having a larger pressure than the thermal gas solution. Moreover, since within the region of the $(T,B)$-phase diagram analyzed in our manuscript the pressure of the plasma increases with $B$ (as also seen on the lattice, see Fig. \ref{fig4}), within this region the black hole solutions are always thermodynamically preferred and do correspond to the true ground state of the system.

As a technical detail, in order to obtain the curves in Fig.\ \ref{fig4} we used a large grid of initial conditions with 720,000 points taking 900 equally spaced points in the $\phi_0$-direction starting from $\phi_0=0.3$ and going up to $\phi_0=7.8$, and 800 equally spaced points in the $\frac{\mathcal{B}}{\mathcal{B}_{\textrm{max}}(\phi_0)}$-direction starting from $\frac{\mathcal{B}}{\mathcal{B}_{\textrm{max}}(\phi_0)}=0$ and going up to $\frac{\mathcal{B}}{\mathcal{B}_{\textrm{max}}(\phi_0)}=0.99$. A large number of points was required to obtain sufficiently smooth curves for $s/T^3$ that allowed for the extraction of the crossover temperature and its dependence on the magnetic field. However, smooth curves for $p$ could be obtained using much smaller (and faster) grids.

\section{Concluding remarks and perspectives}
\label{conclusion}

In this paper we developed, for the first time, a bottom-up holographic model that provides a quantitative description of the crossover behavior observed in the equation of state and in the magnetic susceptibility of a QCD plasma with $(2+1)$-flavors at zero magnetic field. We employed this model to study how an Abelian magnetic field $B$ affects the thermodynamic properties of this strongly coupled plasma (at zero chemical potentials). In the presence of the magnetic field the plasma becomes anisotropic and we used the inflection point of the holographically calculated $s/T^3$ curve to determine how the crossover temperature is affected by the external magnetic field. We found that the crossover temperature decreases with an increasing magnetic field, which agrees with the general behavior recently observed on the lattice. Our model calculations display some level of quantitative agreement with the lattice data for values of the magnetic field up to $eB \lesssim 0.3$ GeV$^2$, which is the expected range achieved in ultrarelativistic heavy ion collisions.

We believe that this agreement with the lattice data can be further improved toward larger values of $eB$ if one tries to carefully match the lattice thermodynamic calculations at $B=0$ by simultaneously taking into account different observables such as the pressure and the speed of sound squared, as we have done in the present approach, with the addition of the entropy density and the trace anomaly in a global fit; in this sense, our choice for the holographic model parameters (fixed at $B=0$) may be systematically improved.

An interesting feature of our holographic model that distinguishes it from other constructions (such as \cite{Ballon-Bayona:2013cta,Mamo:2015dea}) is that the suppression of the crossover temperature with the external magnetic field found here is directly tied to a \emph{quantitative description} of near crossover lattice QCD thermodynamics at $B=0$. It would be desirable to generalize the present holographic model by taking into account the contribution of the chiral condensate. Moreover, motivated by the recent studies in Refs.\ \cite{Endrodi:2015oba,cohen}, one could also investigate if this model indicates the existence of a critical point in the $(T,B)$-plane at higher values of the magnetic field\footnote{Note from Fig. \ref{fig4} that for the values of $B$ considered here we only have a smooth analytical crossover, as also seen on the lattice.}.

The holographic setup constructed here may be employed to obtain estimates for the magnetic field dependence of many other physical observables relevant to the strongly coupled QGP. For instance, one could generalize the calculation of transport coefficients performed in \cite{DK-applications2} and obtain a quantitative estimate of how the anisotropic shear (and bulk) viscosity coefficients vary with the external magnetic field around the QCD crossover transition.

Recently, the effects of an external magnetic field on the equilibration dynamics of strongly coupled plasmas have been studied using holography \cite{DK-applications4,Mamo:2015aia}. In this context, it would be interesting to see how the quasinormal mode spectrum in our nonconformal plasma varies with an external magnetic field. Given that our model can capture the nonconformal behavior of the QGP near the crossover transition, with and without the external magnetic field, a detailed study of the quasinormal modes in this model may shed some light on the thermalization process that takes place in an anisotropic nonconformal strongly magnetized QGP. We hope to report results in this direction in the near future.

\acknowledgments

We thank I.~A.~Shovkovy and E.~S.~Fraga for discussions about the effects of magnetic fields on the QGP and F.~Bruckmann and G.~Endrodi for comments on the manuscript. We also thank the authors of Ref.\ \cite{latticedata3} for making their lattice results avaliable to us. This work was supported by Funda\c c\~ao de Amparo \`a Pesquisa do Estado de S\~ao Paulo (FAPESP) and Conselho Nacional de Desenvolvimento Cient\'ifico e Tecnol\'ogico (CNPq). J.~N.~ thanks the Columbia University Physics Department for its hospitality.

\appendix
\section{Electric susceptibility and conductivity for different coupling functions}
\label{apa}

In order to check further limitations of the present EMD model (some of which have been discussed at the end of Section \ref{sec2.4}), we compare in this Appendix the results for the magnetic susceptibility, and also the electric susceptibility and DC electric conductivity for two different profiles of the Maxwell-Dilaton electric coupling function $f(\phi)$. The first profile is given in Eq. \eqref{2.24}, which was fixed by fitting lattice data \cite{latticedata2} for the magnetic susceptibility at $B=0$, as discussed before. The second profile was fixed in Ref. \cite{Finazzo:2015xwa} by fitting lattice data \cite{Borsanyi:2011sw} for the electric susceptibility also at $B=0$,
\begin{align}
f(\phi)=0.0193\,\textrm{sech}(-100\,\phi)+0.0722\,\textrm{sech}(10^{-7}\,\phi).
\label{A1}
\end{align}

\begin{figure}
\begin{center}
\begin{tabular}{c}
\includegraphics[width=0.45\textwidth]{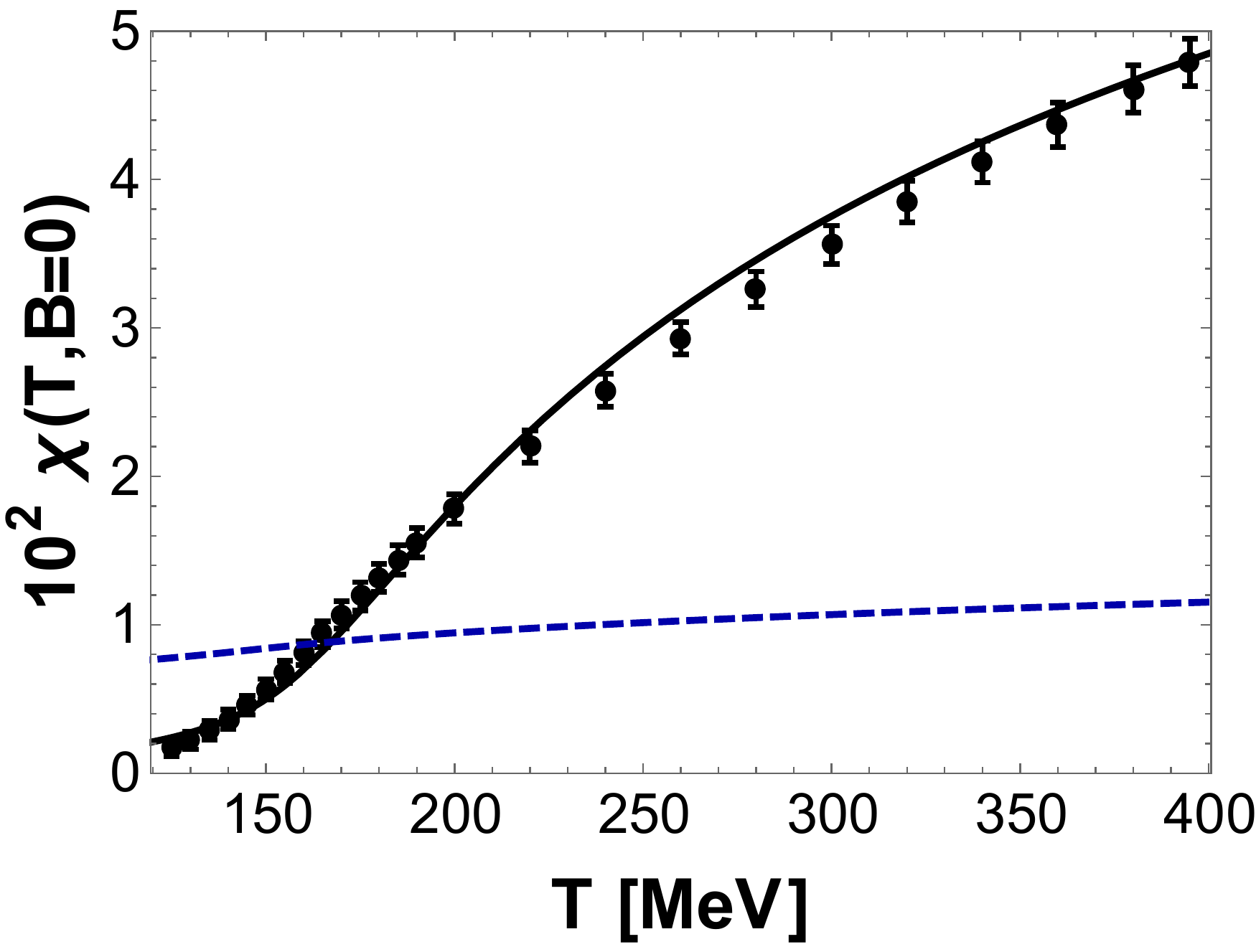} % \\
\end{tabular}
\begin{tabular}{c}
\includegraphics[width=0.45\textwidth]{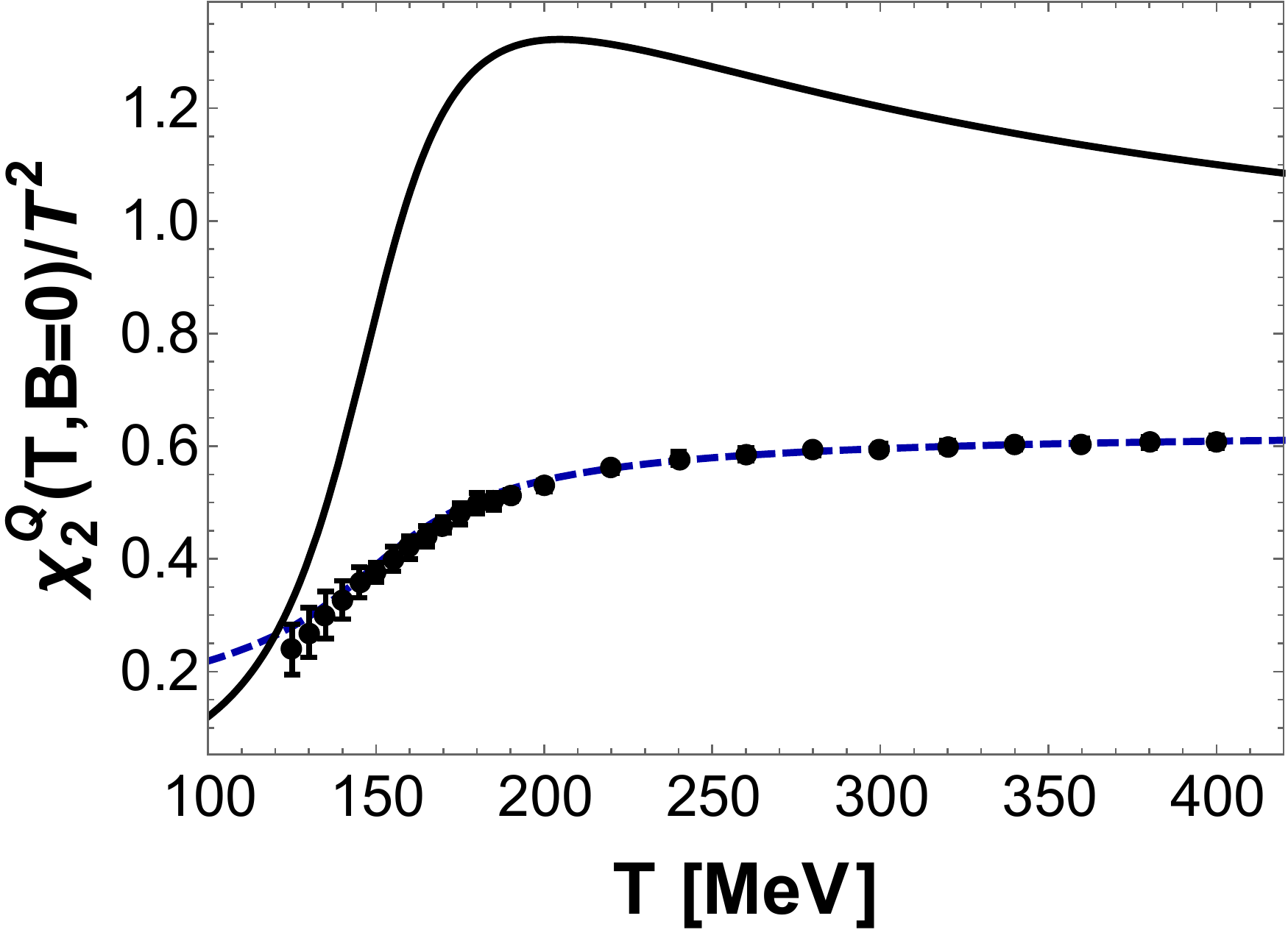} % \\
\end{tabular}
\begin{tabular}{c}
\includegraphics[width=0.45\textwidth]{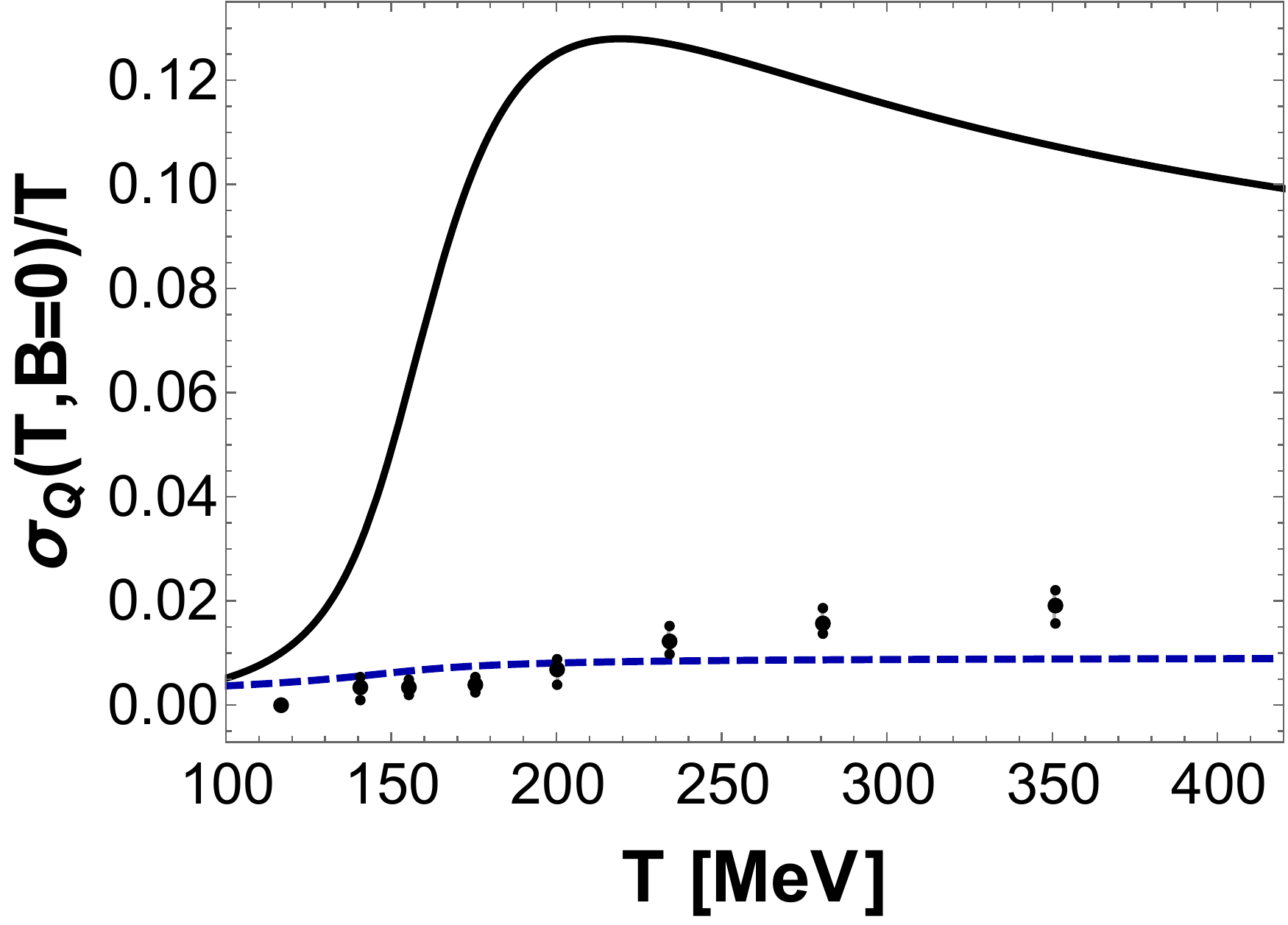} % \\
\end{tabular}
\end{center}
\caption{{\small (Color online) EMD magnetic susceptibility (top left), electric susceptibility (top right) and DC electric conductivity (bottom) for two different choices of the Maxwell-Dilaton electric coupling function $f(\phi)$: the full curves were obtained by using $f(\phi)$ given in Eq. \eqref{2.24}, while the dashed curves were obtained by employing $f(\phi)$ given in Eq. \eqref{A1}. All the lattice data displayed in these plots refer to $(2+1)$-flavor QCD (lattice data for the electric conductivity are taken from \cite{Aarts:2014nba}).}
\label{figlimitations}}
\end{figure}

At $B=0$, the holographic formulas for the electric susceptibility and the DC electric conductivity are given respectively by,
\begin{align}
\frac{\chi_2^Q}{T^2}&=\frac{1}{16\pi^2} \frac{s}{T^3} \frac{1}{f(0)\int_{r_H}^\infty dr\, e^{-2a(r)}f^{-1}(\phi(r))},\label{A2}\\
\frac{\sigma_Q}{T}&=\frac{2\pi\sqrt{h_0^{\textrm{far}}}f(\phi_0)}{\kappa^2},\label{A3}
\end{align}
and we refer the reader to consult Ref. \cite{Finazzo:2015xwa} for a discussion on the derivation of these formulas\footnote{We use that $a(r_H)=a_0=0$ and $\phi(r_H)=\phi_0$.}.

One can see from the results shown in Fig \ref{figlimitations} that a simple EMD holographic model cannot give simultaneously good quantitative descriptions of electric and magnetic field response functions: by adjusting the electric coupling $f(\phi)$ in order to fit the magnetic susceptibility at $B=0$, one is able to attain a good description of the QCD thermodynamics at finite $B$, as shown in Section \ref{sec3}, but response functions to an applied electric field are not well described in a quantitative way within such prescription. On the other hand, if one adjusts the electric coupling $f(\phi)$ in order to match the electric susceptibility, one is not able to obtain a good quantitative agreement with lattice data for the magnetic susceptibility. It would be certainly interesting to think about the construction of some holographic model versatile enough to quantitatively cover the entire electric-magnetic sector of the QGP, which is something that our simple EMD model is not able to do. We must remark, however, that up to now, our EMD model is the only holographic approach available in the literature which is able to match in a quantitative way the behavior of many magnetic field related observables calculated on the lattice.

\end{document}